\documentclass[]{mn2e}
\usepackage{psfig}

\def\mrpd{\hbox{mrad\,d$^{-1}$}}
\def\rpd{\hbox{rad\,d$^{-1}$}}
\def\ps{\hbox{s$^{-1}$}}
\def\rc{\hbox{$r_{\rm c}$}}
\def\mc{\hbox{$m_{\rm c}$}}
\def\omeq{\hbox{$\Omega_{\rm eq}$}}
\def\omsb{\hbox{$\Omega_{\rm sb}$}}
\def\ompo{\hbox{$\Omega_{\rm pole}$}}
\def\dom{\hbox{$d\Omega$}}
\def\dommin{\hbox{$d\Omega_{\rm min}$}}
\def\domvar{\hbox{$d\Omega_{\rm var}$}}
\def\chisq{\hbox{$\chi^2$}}

\def\msun{\hbox{$M_{\odot}$}}

\def\rsun{\hbox{$R_{\odot}$}}

\def\vsin{\hbox{$v \sin i$}}

\def\kms{\hbox{km\,s$^{-1}$}}

\def\degr{\hbox{$^\circ$}}

\begin{document}


\title[Temporal fluctuations in the differential rotation of active stars]
{Temporal fluctuations in the differential rotation of cool active stars} 

\makeatletter

\def\newauthor{%
  \end{author@tabular}\par
  \begin{author@tabular}[t]{@{}l@{}}}
\makeatother
 
\author[J.-F.~Donati et al.]
{\vspace{1.5mm}
J.-F.~Donati$^1$, A.~Collier~Cameron$^2$ and P.~Petit$^1$ \\  
$^1$Laboratoire d'Astrophysique, Observatoire Midi-Pyr\'en\'ees, F--31400 Toulouse, France 
({\tt donati@ast.obs-mip.fr,}\\  \hspace{1mm} {\tt petit@ast.obs-mip.fr})\\
$^2$School of Physics and Astronomy, Univ.\ of St~Andrews, St~Andrews, Scotland KY16 9SS
({\tt acc4@st-andrews.ac.uk})
}

\date{2002, MNRAS, submitted}
\maketitle
 
\begin{abstract} 
This paper reports positive detections of surface differential rotation on two 
rapidly rotating cool stars at several epochs, by using stellar surface 
features (both cool spots and magnetic regions) as tracers of the large scale 
latitudinal shear that distorts the convective envelope in this type of stars.  
We also report definite evidence that this differential rotation is different 
when estimated from cool spots or magnetic regions, and that it undergoes 
temporal fluctuations of potentially large amplitude on a time scale of a few 
years.  

We consider these results as further evidence that the dynamo processes 
operating in these stars are distributed throughout the convective zone rather 
than being confined at its base as in the Sun.  By comparing our observations 
with two very simple models of the differential rotation within the convective 
zone, we obtain evidence that the internal rotation velocity field of the stars 
we investigated is not like that of the Sun, and may resemble that we expect 
for rapid rotators.  We speculate that the changes in differential rotation 
result from the dynamo processes (and from the underlying magnetic cycle) that 
periodically converts magnetic energy into kinetic energy and vice versa.  

We emphasise that the technique outlined in this paper corresponds to the first 
practical method for investigating the large scale rotation velocity field within 
convective zones of cool active stars, and offers several advantages over 
asteroseismology for this particular purpose and this specific stellar class.  
\end{abstract}

\begin{keywords} 
Line: profiles  --
Stars: activity  -- 
Stars: imaging  -- 
Stars: binaries  -- 
Stars: magnetic fields  --
Stars: rotation.  
\end{keywords}

\section{Introduction} 
\label{sect:introduction}

Despite considerable progress obtained over the last century, astrophysicists 
are still far away from a thorough understanding on how magnetic fields are 
produced in the Sun and other cool active stars;  the picture gets even worse 
when it comes to explain quantitatively how these fields generate the 
plethora of active phenomena observed both at their surface (e.g.\ dark 
spots) and in their immediate surrounding (e.g.\ corona, wind).  Particularly 
interesting in this respect is the case of fully convective stars for which 
conventional dynamo processes (supposed to concentrate mostly in the 
interface layer between the radiative interior and the convective envelope, 
i.e.\ the only place where the field can be stored for a time comparable to 
the period of the activity cycle) cannot be invoked to justify, not only the 
presence of a large-scale magnetic field (e.g.\ Johns-Krull \& Valenti 1996), but 
also their very intense activity level.  

The most recent observational results in this field, consisting of temporal 
series of magnetic maps obtained by indirect tomographic imaging from sets of 
rotationally modulated spectropolarimetric data (Donati et al.\ 1992; Donati 
\& Cameron 1997; Donati 1999; Donati et al.\ 1999; Donati et al.\ 2003; 
Petit et al.\ 2003a, 2003b), are particularly intriguing and thus potentially 
very fruitful for providing new clues to what may be happening in the 
convective layers of these stars.  All reconstructed magnetic maps to date 
indeed show the presence of large magnetic regions in the stellar photosphere 
where the field is mostly azimuthal (i.e.\ parallel to the surface and 
oriented along parallels);  moreover, these regions are often observed to 
form partial, or even complete, rings of azimuthal field encircling the star 
at various latitudes, and are thus interpreted as the direct detection, at 
photospheric level, of the quasi-axisymmetric toroidal component of the 
large-scale dynamo field.  

These results suggest in particular that dynamos operating in very active 
stars are likely to be significantly different than that of the Sun.  The 
observations indeed reveal that such dynamos are able to produce large scale 
field (and especially toroidal field) very close to the surface (since we 
would not observe toroidal field in the photosphere otherwise), and 
presumably even throughout the whole convective zone;  dynamo processes in 
these stars are therefore likely distributed within the convective envelope, 
rather than being confined at its base as in the Sun.  This conclusion 
raises however a number of subsequent, and yet unresolved, problems;  we can 
no longer understand, for instance, how the magnetic field can be stored 
within the convective zone for periods of time as long as decades.  At the 
same time, it may provide new insight for a number of long standing mysteries, 
and in particular for our comprehension on how dynamos operate in fully 
convective stars.   

One of these mysteries concerns the long-term fluctuations observed in the 
orbital period of close binary stars in which one member is at least a 
cool active star (e.g.\ Hall 1990), and occurring on a time scale of a few 
decades, i.e.\ much shorter than what would be required for tidal coupling 
between system components to operate (Zahn 1989).  The least unreasonable 
interpretation for this phenomenon remains that proposed by Applegate (1992); 
he suggests that dynamo processes at work in one star of the binary system activate a 
periodic exchange between magnetic and kinetic energy within the convective 
zone of this star, and thus a cyclic fluctuation of its quadrupolar moment as 
well as of its gravitational field.  Given the amplitude of the observed 
fluctuations in orbital period, one can easily infer that such magnetic to 
kinetic energy exchanges specific to dynamo processes must occur throughout 
the whole convective zone and not only within a thin layer (Lanza et al.\ 
1998; Donati 1999), i.e.\ that dynamo processes in these stars should indeed 
be distributed.   

If Applegate's (1992) idea is true, it therefore implies that very active 
stars such as those found in close binary stars, but also young single stars 
that did not dissipate most of their angular momentum yet and still exhibit very 
energetic active phenomena, should feature a convective zone that globally 
undergoes such periodic exchanges between magnetic and kinetic energy.  It 
means in particular that the internal velocity field within the convective 
zone of very active stars, and therefore both their radial and surface 
differential rotation profiles, should vary with time in a way correlated 
to the magnetic cycle.  Detecting such variations in the differential 
rotation of active stars thus appears as a very interesting observational 
challenge, as it would provide a definite confirmation of both Applegate's 
(1992) mechanism and of the existence of distributed dynamos.  It would also 
bring into a much wider astrophysical context the recent discovery that the 
solar angular rotation at the base of the convective zone is undergoing 
temporal fluctuations (Howe et al.\ 2000) as a probable consequence of the 
activity cycle.  

Measuring surface differential rotation of very active stars is however a 
rather tricky task;  detecting changes of this differential rotation is 
even more difficult.  Several methods have been proposed to estimate surface 
differential rotation.  Some of them try to make use of the subtle changes 
that differential rotation induce in the profile of spectral lines (Brunning 
1981; Reiners \& Schmitt 2002);  this method can however not be used for the 
stars we are interested in, the spectral lines of which being heavily 
distorted by the presence of cool spots at their surfaces.  Other methods 
(e.g.\ Donati \& Cameron 1997) propose to take advantage of these surface 
features, and use them as tracers to derive some information on how their 
rotation periods depend on latitude.  The most recent of such techniques, 
that of Cameron et al.\ (2002), succeeded, not only in estimating surface 
differential rotation of one active star, but also in suggesting that 
temporal changes in the amount of differential rotation were indeed occurring 
in this star (Cameron \& Donati 2002).  

In this paper, we propose to make use of yet another technique, first proposed 
by Donati et al.\ (2000) to estimate the differential rotation of a young 
pre-main-sequence star, then by Petit et al.\ (2002, 2003a, b) both for 
simulation purposes and application to spectropolarimetric data.  The aim of 
this paper is to apply this method to the extensive spectropolarimetric data 
set that we collected in the last seven yr for three stars, namely the young 
ultra-fast rotator AB~Doradus, the young K0 star LQ~Hydrae and the K1 subgiant 
of the RS~CVn system HR~1099, and from which yearly brightness and magnetic 
maps were obtained and published in the literature (Donati \& Cameron 1997; Donati 1999; Donati 
et al.\ 1999, 2003).  In Sect.~\ref{sect:modelling}, we recall the main 
aspects of both observational material and modelling tool, then describe 
its application to the three selected stars in Sect.~\ref{sect:diffrot};  
after discussing at length the implication of our results for the 
understanding of the global dynamics of convective zones of cool stars (in 
Sect.~\ref{sect:discussion}), we finally conclude and propose in 
Sect.~\ref{sect:conclusion} a few directions in which this work could be 
fruitfully extended.

\section{Observational material and modelling techniques} 
\label{sect:modelling}

The data we use are the spectropolarimetric observations collected at the 
Anglo-Australian Telescope (AAT) with a visitor polarimeter mounted at 
Cassegrain focus and fibre linked to the high resolution UCL Echelle 
Spectrograph (UCLES).  This material, and in particular the observing logs, 
the observing procedures and the data reduction details, are described 
extensively in a series of published papers that present the results obtained 
up to now (Donati et al.\ 1997; Donati \& Cameron 1997; Donati 1999; Donati et 
al.\ 1999, 2003).  

\begin{figure*}
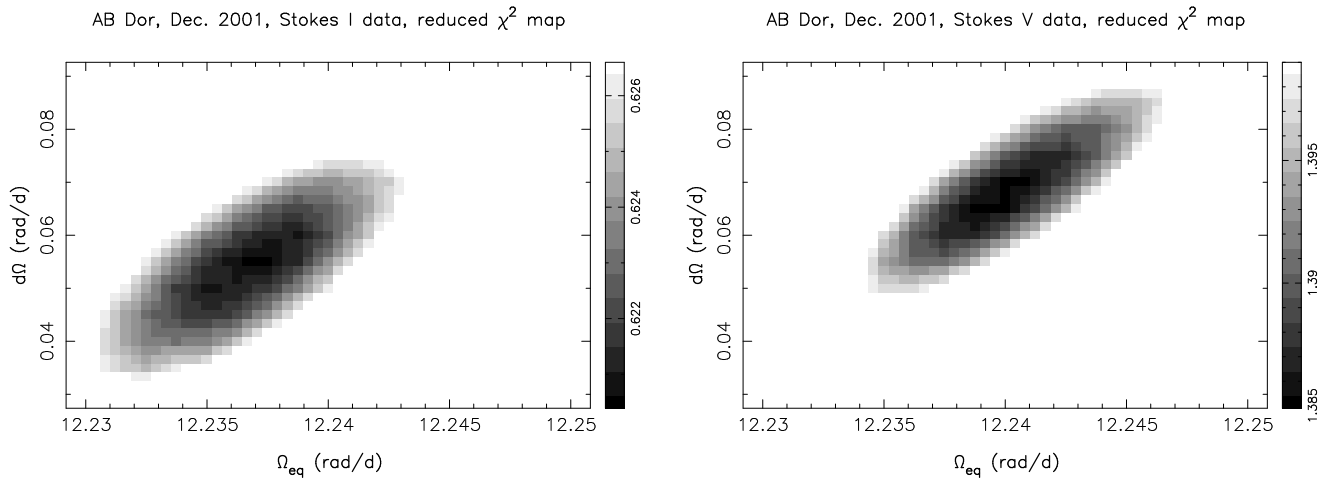

\flushleft{\hbox{\psfig{file=fig/evol_csqi.ps,angle=270,width=8.4cm}
           \hspace{5mm}\psfig{file=fig/evol_csqv.ps,angle=270,width=8.4cm}}} 
\caption[]{Reduced \chisq\ maps in the differential rotation parameter plane, obtained for AB~Dor 
from the Stokes $I$ (left panel) and Stokes $V$ (right panel) data sets collected at epoch 2001.99.  
Black corresponds to models for which the minimum \chisq\ level was obtained (given the assumed 
reconstructed information content), while white depicts those for which the \chisq\ is larger by  
at least 1\% (i.e.\ by least 274 and 69 for the left and right panels respectively) than the 
minimum \chisq.  The outer envelope of all grey points corresponds to 11 and 5.5~$\sigma$ 
confidence ellipses for both differential rotation parameters (for the left and right panels 
respectively), while its projection on each axis gives the 16 and 8~$\sigma$ confidence intervals 
for each parameter taken separately.  Note that distances on both axes are not equal.  }
\label{fig:csqmaps}
\end{figure*}

The data we use in this paper are all observations of AB~Dor, LQ~Hya and 
HR~1099 collected since epoch 1995.9, i.e.\ obtained in a similar spectrograph 
configuration.  This makes a maximum amount of 7 data sets (corresponding to 
epochs 1995.94, 1996.99, 1998.03, 1999.00, 1999.97, 2000.93 and 2001.99) for 
each of the three stars;  we however anticipate that several of them (and in 
particular those with very few data or collected over a timespan of no more 
than a few days, e.g.\ those at epoch 1999.97) will be unusable for our 
purpose, according to the conclusions of Petit et al.\ (2002).  
Note that all data were reprocessed with the newest version 
of the reduction software (Donati et al.\ 2003) to ensure the highest possible 
internal accuracy in radial velocity as well as the best level of homogeneity.  
All data were run through Least-Squares Deconvolution (LSD), a multiline tool 
developed by Donati et al.\ (1997) found to extract successfully the average 
rotationally broadened shape of unpolarised or circularly polarised lines 
(called Stokes $I$ and $V$ LSD profiles in the following) at each observed 
phase of the stellar rotational cycle, from the thousands of moderate to 
strong spectral features present in the recorded wavelength domain.  

As mentioned earlier, the method we chose to estimate surface differential 
rotation for the selected stars is that introduced by Donati et al.\ (2000), 
the validity of which was confirmed through an extensive sets of simulations 
by Petit et al.\ (2002).  Recently, Petit et al.\ (2003a, b) applied it 
successfully to large data sets collected over several years on a different 
telescope and with a different (though similar) instrumentation for two different 
stars, one of them being our third object of interest (the K1 subgiant of 
HR~1099) and the other a giant star of FK~Com type (HD~199178).  
This method consists in assuming a given surface differential 
rotation law (that we implement in our indirect stellar surface imaging code) 
and producing brightness or magnetic images of the stellar surface from all 
unpolarised or circularly polarised data available for this star at a given 
season;  we then take as most probable the differential rotation law that 
generates the images with lowest information content (given a certain data fit 
accuracy), or equivalently the images that provide the best fit to the data 
(given a certain image information content).  In practice, both options yield 
very similar results;  the second one happens to be more convenient as it 
provides an easy way of estimating error bars on the differential rotation 
parameters (see Petit et al.\ 2002).  

The stellar surface imaging code that we use is that of Brown et al.\ (1991) 
and Donati \& Brown (1997), applied quite extensively to real data in the last 
decade (the latest examples being Donati et al.\ 2003 and Petit et al.\ 2003a, 
b).  It includes in particular the possibility of reconstructing images of 
stars that are differentially rotating, i.e.\ of taking into account 
explicitly the differential rotation pattern in the imaging process.  
The surface differential rotation law we assume in this paper is given by 
the following equation: 
\begin{equation}
\label{eq:diffrot}
\Omega(\theta) = \omeq - \dom \cos^2 \theta
\end{equation}
\noindent where $\Omega(\theta)$ is the rotation rate at colatitude $\theta$, 
\omeq\ the rotation 
rate of the equator and \dom\ the difference in rotation rate between the pole 
and equator.  At each season, one brightness and one magnetic image is 
produced from the full Stokes $I$ and Stokes $V$ data sets respectively, for 
each selected pair of differential rotation parameters within a reasonable 
search domain of the \omeq--\dom\ plane.  This typically means that we have to 
compute a few hundred brightness and magnetic images for each star at each 
season, which therefore requires large amount of computing time;  for instance, 
in the particular case of the Stokes $I$ data set of AB~Dor at epoch 2001.99 
where as much as 27,400 data points must be fitted simultaneously, we typically 
need a few days of computing time on the fastest available workstations to derive 
how the reduced \chisq\ of the fit to the data varies with the differential rotation parameters 
at a good enough resolution.  A typical example of the reduced \chisq\ maps we 
obtain from Stokes $I$ and $V$ data is shown in Fig.~\ref{fig:csqmaps}.  

It could be argued that, in the sake of computing time, we could actually 
derive the optimal differential rotation parameters as part of the imaging 
process itself, since this would mean no more than fitting another two 
parameters in addition to the thousands of image pixels we already have to deal 
with.  However, our experience is that the \chisq\ maps we obtain are not 
always as nice as those shown in Fig.~\ref{fig:csqmaps}, and sometimes do not 
feature any physically meaningful minimum.  We therefore preferred to keep our 
initial procedure, and check in each case whether the obtained reduced \chisq\ 
maps showed a clear enough minimum to be able to claim safely that differential 
rotation is indeed detected at the corresponding epoch.  Another very important 
advantage of producing reduced \chisq\ maps is that they can also be used in a 
very straightforward way to obtain error bars on our estimates of the 
differential rotation parameters.  Following Press et al.\ (1992), we indeed 
know that the 1~$\sigma$ confidence interval for each differential rotation 
parameter taken separately can be obtained by searching for all points of the \omeq--\dom\ plane 
for which the \chisq\ increase (with respect to the minimum value in the map) 
is smaller than 1\footnote{These error bars should not be confused with the error 
ellipses for both parameters taken simultaneously, such as those depicted in 
Figs.~\ref{fig:abdriv} to \ref{fig:hrdriv}, which includes all points of the map 
for which the \chisq\ increase is smaller than 2.3} 
(or equivalently for which the reduced \chisq\ increase is 
smaller than $1/n$, where $n$ is the number of fitted data points), and 
projecting this ellipsoid onto the \omeq\ and \dom\ axes respectively.  The 
simplest way to evaluate these projections (and thus the formal error bars 
for each parameter) is to compute the curvature radii of the \chisq\ paraboloid 
at its minimum and the correlation coefficient between the two parameters;  
for a \chisq\ paraboloid with linear and quadratic terms given by 
$a\ \omeq^2 + b\ \omeq\ \dom + c\ \dom^2 + d\ \omeq\ + e\ \dom$, the 
value of the optimal parameters and the associated error bars simply write: 
\begin{equation}
\label{eq:eqomeq}
\omeq = (b e - 2 c d) / (4 a c - b^2),\hspace{2mm} \sigma (\omeq) = (a-b^2/4 c)^{-1/2} 
\end{equation}
\begin{equation}
\label{eq:eqdom}
\dom  = (b d - 2 a e) / (4 a c - b^2),\hspace{2mm} \sigma (\dom)  = (c-b^2/4 a)^{-1/2}
\end{equation}

\begin{table*}
\caption[]{Summary of differential rotation parameters obtained for AB~Dor on each season.  
Columns~2 to 6 correspond to Stokes $I$ data while columns~7 to 11 to Stokes $V$ data. 
Cols.~2 and 7 list the derived equatorial rotation rate \omeq\ with its 68\% (i.e.\ 1~$\sigma$) 
confidence interval, Cols.~3 and 8 the difference in rotation rate \dom\ 
between the equator and pole 
with its 68\% confidence interval, Cols.~4 and 9 gives the inverse slope of the ellipsoid in the 
\omeq--\dom\ plane (also equal to $\cos^2 \theta_s$ where $\theta_s$ denotes the colatitude 
of the gravity centre of the spot distribution, see Donati et al.\ 2000), Cols.~5 and 10 the 
rotation rate $\Omega_s$ at colatitude $\theta_s$, and Cols.~6 and 11 the total number of 
data points used in the imaging process.   }
\begin{tabular}{ccccccccccc}
\hline
        & \multicolumn{5}{c}{Stokes $I$ data} & \multicolumn{5}{c}{Stokes $V$ data} \\
 &&&&&&&&&& \\
Epoch   & \omeq\ & \dom\ & $\cos^2 \theta_s$ & $\Omega_s$ & $n$ & \omeq\ & \dom\ & $\cos^2 \theta_s$ & $\Omega_s$ & $n$ \\
        & (\mrpd) & (\mrpd) & & (\rpd) & & (\mrpd) & (\mrpd) & & (\rpd) & \\
\hline 
1995.94 & $12,242.1\pm0.8$ & $53.4\pm2.5$ & 0.277 & 12.227 & 13,398 & $12,253.8\pm3.1$ & $96.7\pm8.1$ & 0.356 & 12.219 & 3,498 \\
1996.99 & $12,235.9\pm0.8$ & $47.1\pm2.5$ & 0.277 & 12.223 & 23,958 & $12,242.8\pm1.1$ & $59.6\pm3.1$ & 0.301 & 12.225 & 5,940 \\
1999.00 & $12,239.1\pm0.4$ & $58.4\pm1.5$ & 0.157 & 12.230 & 21,600 & $12,251.3\pm3.2$ & $79.5\pm7.4$ & 0.412 & 12.219 & 5,300 \\
2000.93 & $12,235.1\pm0.9$ & $46.1\pm2.8$ & 0.281 & 12.222 & 14,000 & $12,242.6\pm2.1$ & $74.1\pm6.0$ & 0.272 & 12.222 & 3,500 \\
2001.99 & $12,236.6\pm0.4$ & $54.0\pm1.3$ & 0.221 & 12.225 & 27,378 & $12,240.1\pm0.8$ & $68.2\pm2.4$ & 0.262 & 12.222 & 6,903 \\
\hline 
\end{tabular}
\label{tab:abdor}
\end{table*}

The procedure we thus carry out to obtain both the optimal differential rotation 
parameters and associated error bars simply consists in fitting the reduced 
\chisq\ maps (actually only the points in the neighbourhood of the minimum) 
by a bi-dimensional paraboloid (as expressed above), derive the five coefficients 
of this paraboloid and then the requested quantities using  
Eqs.~\ref{eq:eqomeq} and \ref{eq:eqdom}.  In addition to being the easiest way to 
evaluate both the position of the minimum and the curvature radii at that 
minimum, it also averages out the slight random differences in convergence 
accuracy that always remain at a relative level of 0.01\% between all points 
in the map.  We find this method to be very robust in practice, and the 
derived differential rotation parameters very weakly sensitive to internal 
parameters, such as for instance the amount of image information towards which 
the code is asked to converge.

\section{Surface differential rotation} 
\label{sect:diffrot}

As explained in Petit et al.\ (2002), estimating the surface differential 
rotation of a cool spotted star requires that the spectroscopic (or 
spectropolarimetric) data set used in this aim correctly samples in time the 
variability in the surface distribution of cool spots or magnetic regions 
associated with this phenomenon.  It means in particular that large fractions 
of the stellar surface must be observed several times, with adequate temporal 
gaps in between.  It straightforwardly tells us that sparse data sets, such as 
those obtained for all three stars in 1998 January (epoch 1998.03) and 1999 
December (epoch 1999.97) in which most regions of the stellar surfaces was 
observed only once, are not be appropriate for our purposes.  Similarly, the 
data set collected on LQ~Hya at epoch 1995.94, containing only two main 
groups of observations (plus a third one of very poor quality, Donati et al.\ 
1997; Donati 1999), is totally useless for the present investigation.  

\subsection{The young ultra-fast rotator AB~Doradus} 
\label{sect:abdor}

As far as AB~Dor is concerned, it leaves us with five main data sets 
(corresponding to epochs 1995.94, 1996.99, 1999.00, 2000.93 and 2001.99) with 
adequate phase overlap between different groups of observations.  The main 
reason for this is of course that the rotation period of AB~Dor is short 
enough (0.51479~d, equivalent to a rotation rate of 12.2053~\rpd) to allow a 
large fraction of the stellar surface (as much as two thirds) to be observed 
in a single night, and sufficiently close to half a day that reasonable phase 
overlap is obtained between data sets separated by a few days.  The estimates 
we obtain are listed in Table~\ref{tab:abdor} and shown visually on 
Fig.~\ref{fig:abdriv}.  The imaging parameters that we assumed for AB~Dor in 
this experiment are those derived by Donati et al.\ (2003), with the 
inclination of the rotation axis to the line of sight $i$ set to 60\degr\ and 
the projected equatorial rotation velocity \vsin\ set to 89~\kms.  

\begin{figure*}
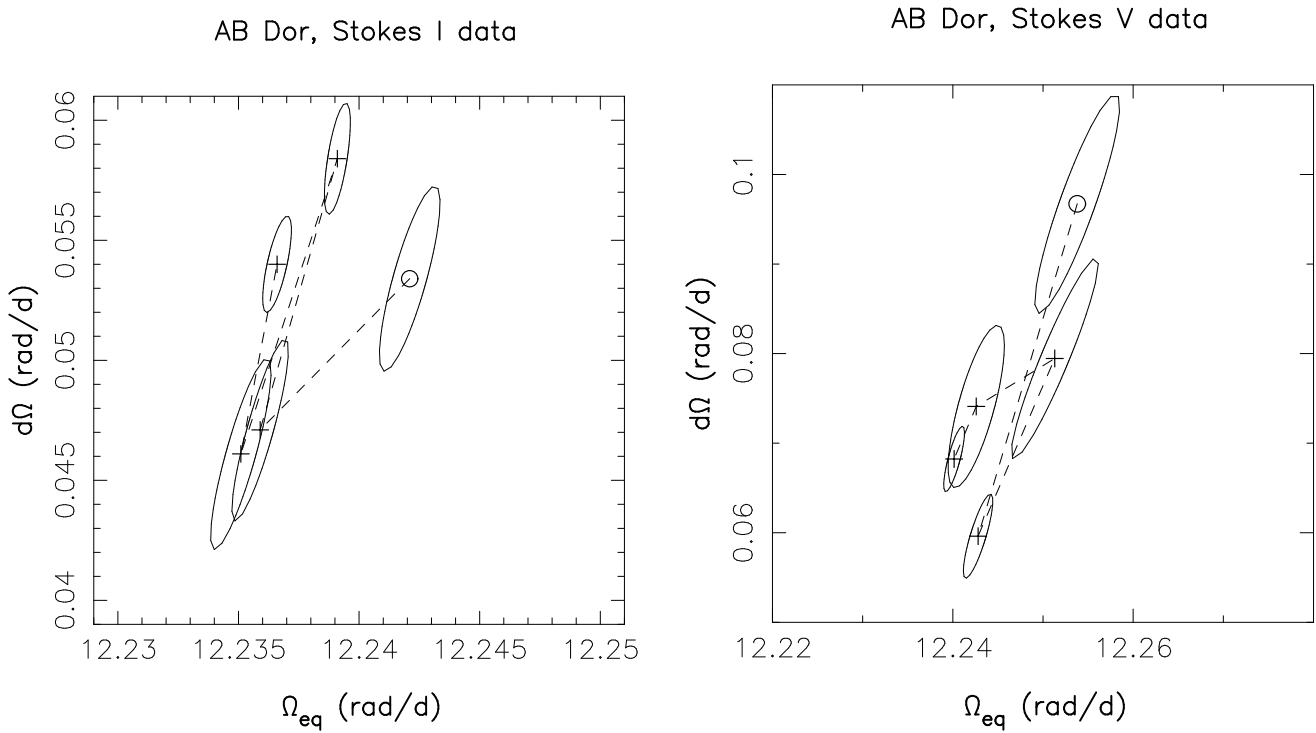

\flushleft{\hbox{\psfig{file=fig/evol_abdri.ps,angle=270,width=8.4cm}
           \hspace{5mm}\psfig{file=fig/evol_abdrv.ps,angle=270,width=8.4cm}}} 
\caption[]{Differential rotation parameters of AB~Dor as measured from Stokes 
$I$ (left panel) and Stokes $V$ data (right panel).  All points are marked with 
$+$ signs and joined together with a dashed line by order of increasing epoch 
starting from the first point in the series (corresponding to epoch 1995.94 and 
marked as $\circ$).  For each point, 68\% confidence ellipses for both parameters 
(projecting on both axes into 1.5~$\sigma$ confidence intervals for each parameter 
taken separately) are also depicted.  Note the different scale used for each plot. } 
\label{fig:abdriv}
\end{figure*}

Several comments can be made from these results.  First of all, we want to 
stress the overall good agreement of the estimates obtained from either the 
Stokes $I$ or the Stokes $V$ data.  If the subtle signature we are attempting 
to retrieve from our spectra was not produced by surface differential rotation, 
but rather by some spurious problem in the analysis, then there would be no 
reason for us to find roughly the same result five times, from five totally 
independent data sets.  It represents already a strong argument that what we 
see is indeed surface differential rotation.  We can then also note that the 
values we derive from Stokes $I$ data at epoch 1995.94 are in good agreement 
(i.e.\ within about one error bar) with the estimates we originally published 
($\omeq=12.2434$~\rpd, $\dom=56.4$~\mrpd) when differential rotation was first 
detected on AB~Dor (Donati \& Cameron 1997), indicating that the new method 
we are using is well behaved and provides results that are compatible with 
those yielded by older and more straightforward (though less accurate and 
rigorous) techniques.  

There is however 
a number of significant differences between the results obtained recently with 
yet another technique (estimating the recurrence rate and latitude of 
individual starspots by accurately tracing the repetitive trails they produce 
in dynamic spectra each time they cross the visible stellar hemisphere) and 
published by Cameron \& Donati (2002).  For the two epochs (actually the two 
data sets) we have in common, these authors obtain differential rotation 
parameters of $\omeq=12.2498\pm0.0019$~\rpd\ and $\dom=71.4\pm5.8$~\mrpd\ for 
epoch 1995.94, and $\omeq=12.2417\pm0.0017$~\rpd\ and $\dom=57.6\pm4.8$~\mrpd\  
for epoch 1996.99, i.e.\ way out both our and their sets of errors bars.  Note 
that the variation of both parameters between the first and second epochs is 
found to be roughly similar in both studies, suggesting that most of the 
discrepancy is likely attributable to slight internal calibration errors in 
the second method but does not question the reality of the measurements 
themselves.  We indeed find that this second method is much more sensitive 
than the one used here to the assumed \vsin\ and that part (about half) of the 
discrepancy we report comes from the different \vsin\ used in both studies, set 
to 89~\kms\ for our experiment (as obtained by Donati et al.\ 2003) and to 
91~\kms\ in Cameron \& Donati (2002).  The remaining difference seems to be 
attributable to the way Cameron \& Donati (2002) weigh the rotation rates 
derived from individual spots when it comes to fitting them with a differential 
rotation law;  if, instead of using roughly equal weight for all spots as they 
did, larger weight is attributed to bigger spots (as our technique implicitly 
does), a much better (though still not quite perfect) agreement is obtained.  
It is not clear yet whether this problem reflects no more than numerical 
uncertainties, or something intrinsic to the star, e.g.\ that small and large 
spots do not happen to suffer the same latitudinal shear (as is the case on the 
Sun).  

Although the different results obtained from either Stokes $I$ and $V$ data are 
roughly all of the same magnitude, they nevertheless show definite differences 
among them, as one can see quite straightforwardly from Fig.~\ref{fig:abdriv}.  
From Stokes $I$ data for instance, we note that \omeq\ changed by as much as 
6.2~\mrpd\ between epoch 1995.94 and 1996.99, about 7.8 times the quoted error 
bar;  comparing now how \dom\ evolved between epoch 1996.99 and 1999.00, we 
obtain a variation of 11.2~\mrpd, about 4.5 times the largest error bar.  
Similarly, results from Stokes $V$ data indicate variations as large as 
13.7 and 37.1~\mrpd\ in \omeq\ and \dom\ respectively, again about 4.5 times 
the quoted error bars (of 3.1 and 8.1~\mrpd\ respectively).  This result by 
itself is enough to safely claim that we indeed detected fluctuations in the 
differential rotation parameters of AB~Dor (with a false alarm probability 
much smaller than 1~ppm), thus confirming the initial report of Cameron \& 
Donati (2002).  Our conclusion relies of course on the fact that our error bars 
are not significantly underestimated;  although the simulations of Petit et 
al.\ (2002) reveal that this may indeed happen when the phase sampling of the 
observations is sparse, they nevertheless establish quite firmly that this 
does not occur with as dense a sampling as that secured for AB~Dor.  

\begin{table*}
\caption[]{Same as Table~\ref{tab:abdor} for LQ~Hya}
\begin{tabular}{ccccccccccc}
\hline
        & \multicolumn{5}{c}{Stokes $I$ data} & \multicolumn{5}{c}{Stokes $V$ data} \\
 &&&&&&&&&& \\
Epoch   & \omeq\ & \dom\ & $\cos^2 \theta_s$ & $\Omega_s$ & $n$ & \omeq\ & \dom\ & $\cos^2 \theta_s$ & $\Omega_s$ & $n$ \\
        & (\mrpd) & (\mrpd) & & (\rpd) & & (\mrpd) & (\mrpd) & & (\rpd) & \\
\hline 
1996.99 &                 &              &       &       &       & $3,892.3\pm9.6$ &$-48.7\pm23.5$& 0.355 & 3.910 &   299 \\
2000.93 & $3,982.4\pm9.1$ &$194.2\pm21.6$& 0.398 & 3.905 & 2,835 & $3,969.3\pm22.9$&$200.9\pm72.4$& 0.298 & 3.909 &   805 \\
2001.99 & $3,922.3\pm1.0$ & $14.4\pm2.9$ & 0.333 & 3.918 & 5,945 & $3,944.6\pm2.3$ & $86.7\pm8.7$ & 0.213 & 3.926 & 1,558 \\
\hline 
\end{tabular}
\label{tab:lqhya}
\end{table*}

A rather surprising discovery is that the differential rotation parameters 
obtained from Stokes $V$ data are clearly larger, by an amount that varies 
between 25\% to as much as 80\%, than those derived from Stokes $I$ data, for 
a reason that we do not fully understand yet.  The effect 
in itself is already clearly visible on the raw \chisq\ maps themselves, such 
as those shown in Fig.~\ref{fig:csqmaps} for instance.  One can in particular 
wonder whether it is not due to some artifact inherent to the method itself.  
The first natural suspicion is that the surface differential rotation law we 
assumed in the imaging process is significantly different from the true one;  
in this case, potential differences in the latitude range respectively sampled 
by the brightness and magnetic regions recovered at a given epoch may explain 
the observed discrepancy between values of \omeq\ and \dom\ derived from Stokes 
$I$ and $V$ data.  If this was the correct interpretation however, we would 
then expect to see larger discrepancies when the latitudinal barycentres of the 
brightness and magnetic distributions are further away from one another.  
This is certainly not what we get, e.g.\ at epoch 2000.93 where both the 
latitudinal barycentres of the brightness and magnetic distributions almost 
coincide (see Table~\ref{tab:abdor}) while the photospheric shear derived 
from Stokes $V$ data exceeds that deduced from Stokes $I$ data by more than 
60\%.  We thus have to conclude that the origin of the observed discrepancy 
is somewhere else.   

Another potential explanation may be looked for in the fact that 
magnetic features hosting radial field 
lines, suffering from a stronger dependence to limb darkening than brightness 
features (Brown et al.\ 1991), are reconstructed with a stronger bias towards 
higher latitudes than cool spots;  matching the observed recurrence rate of 
surface features would then require larger \omeq\ and \dom\ when magnetic 
regions are considered rather than cool spots.  This however cannot be the 
case.  The differential latitude bias that we mention remains indeed rather 
small, especially when phase coverage is as dense as that we have on most 
seasons, and does certainly not exceed a few degrees.  Moreover, observations 
indicate that magnetic field is, most of the time, predominantly azimuthal 
(Donati et al.\ 2003) rather than radial, and azimuthal field regions are 
actually not sensitive to the above mentioned effect.  We can thus safely 
claim that the considerably larger shear observed from Stokes $V$ data cannot 
be explained this way.  Again, the best method for investigating whether this 
phenomenon is genuine or spurious is to carry out simulations;  this is what 
Petit et al.\ (2002) did, leading to the conclusion that no such bias was to 
be observed between differential rotation parameters estimated from either 
Stokes $I$ or Stokes $V$ data when phase sampling was as dense as ours.  
Similar simulations, either on fake data or on the real observations 
presented here, also enable us to conclude that reasonable errors in the 
imaging parameters (for instance errors of the order of 2~\kms\ in \vsin) 
produce only small variations in the results (of the order of one error bar 
typically) and in particular negligible changes in this discrepancy we 
report between differential rotation parameters obtained from Stokes $I$ 
and $V$ data.  

The only conclusion we are left with is that the observed difference reflects 
some genuine property of surface features in cool active stars, for instance 
that magnetic regions are not anchored at the same depth in the convective zone
and thus do not witness the same shear.  Direct evidence from this can be found 
by comparing the average latitude at which cool spots and magnetic regions are 
found to appear, as well the estimated recurrence rate at this average latitude 
(both parameters being listed in Table~\ref{tab:abdor}).  At epoch 1999.00 for 
instance, we obtain that magnetic regions preferentially form at a latitude of 
39.9\degr\ and rotate at an average rate of 12.219~\rpd, while cool spots tend to show up 
at a latitude 23.3\degr\ and rotate at a mean rate of 12.230~\rpd;  the latitudinal 
shear we derive from this, equal to 43~\mrpd, is significantly smaller than 
that obtained by using either Stokes $I$ data alone ($58.4\pm1.5$~\mrpd) or 
Stokes $V$ data only ($79.5\pm7.4$~\mrpd).  It clearly indicates that both sets 
of features do not refer to the same latitudinal angular velocity field, and 
therefore likely not to the same depth in the convective zone.  We will come 
back on this point in Sect.~\ref{sect:discussion}.  Note as well that the 
pattern of temporal variation in the differential rotation parameters is not 
the same either between Stokes $I$ and Stokes $V$ data;  looking at each change 
from one season to the next in the \omeq--\dom\ plane as a vector of given 
length and direction, we observe that the vectors associated with Stokes $I$ data 
do not necessarily feature the same direction as (and sometimes even point in 
the opposite direction to) those associated with Stokes $V$ data (the most 
obvious example being the evolution between epoch 2000.93 and 2001.99).  Again, 
we consider this as additional evidence that both sets of features correspond 
to different zones of the convective envelope.  

Note that this difference in differential rotation as revealed by cool spots 
and magnetic features was not seen in the initial study of Donati \& Cameron 
(1997), probably because only one component of the reconstructed magnetic field 
(the radial field map) was used at that time in the cross correlation 
experiment from which the photospheric shear was measured;  in the present 
study, we make use of all available information in Stokes $V$ profiles, 
implying that the present results are definitely more reliable.  

\subsection{The young K0 dwarf LQ~Hydrae} 
\label{sect:lqhya}

For LQ~Hya, only three data sets (those corresponding to epochs 1996.99, 
2000.93 and 2001.99) happened to be rich enough to yield a positive detection 
of surface differential rotation.  The one obtained at epoch 1999.0, although 
featuring a rather large number of observations (54 profiles in Stokes $I$ and 
14 in Stokes $V$) yields \chisq\ maps with no clearly defined paraboloid, 
most likely as a result of the much poorer quality of the data secured during 
this campaign (see Donati et al.\ 2003).  The opposite situation is observed 
with the data collected at epoch 1996.99;  although gathered in a period of 
only 5~d, they are nevertheless able to provide us with a meaningful estimate 
of surface differential rotation, but only from the Stokes $V$ subset.  The 
results we obtain are listed in Table~\ref{tab:lqhya} and shown in 
Fig.~\ref{fig:lqdriv}.  The imaging parameters we use for LQ~Hya are again 
those determined by Donati et al.\ (2003), equal to $i=60\degr$ and 
$\vsin=26$~\kms.   

\begin{figure}
\flushleft{\psfig{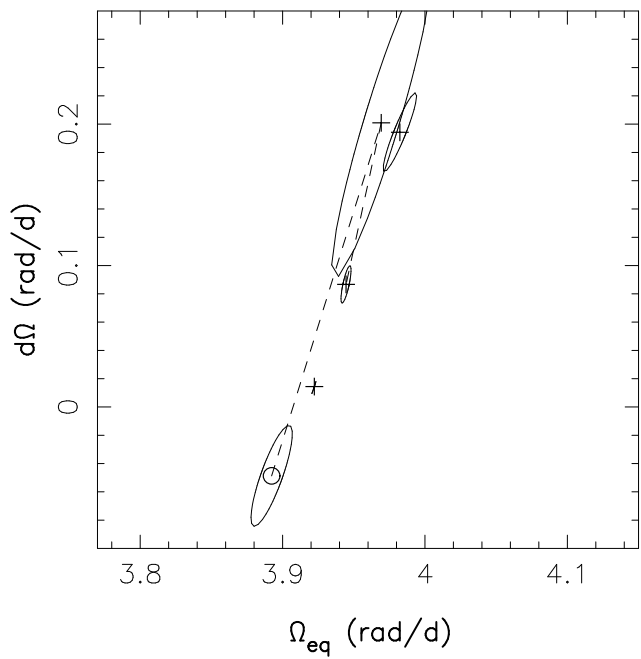}}
\caption[]{Differential rotation parameters of LQ~Hya.  The dashed line links 
together measurements from Stokes $V$ data  by order of increasing epoch 
(starting from epoch 1996.99).  As in Fig.~\ref{fig:abdriv}, 
68\% confidence ellipses for both parameters are also shown.  }
\label{fig:lqdriv}
\end{figure}

The first thing worth reporting is of course the detection of differential 
rotation itself at the surface of LQ~Hya;  although this discovery is not 
altogether very surprising (since we probably expect all stars with a 
significant convective envelope to rotate differentially), it is nevertheless 
the first detection of differential rotation on this star and the second one 
on a zero-age-main-sequence star (after AB~Dor), making it in this respect 
very interesting for potential comparisons with the results obtained on AB~Dor.  
In that respect, the very large photospheric shear detected at epoch 2000.93 
(equal to about 0.2~\rpd, implying a time for the equator to lap the pole by 
one complete turn of only one month) is about $3\pm1$ times larger than that of the 
Sun or than the average \dom\ we measure for AB~Dor.  Given the fact that 
LQ~Hya and AB~Dor are very similar in terms of internal structure (Donati et 
al.\ 2003), it already clearly demonstrates that any relation we may eventually 
derive from the observations and giving the strength of this shear as a 
function of, say, stellar mass (e.g.\ Kitchatinov \& R\"udiger 1999) is 
expected to show a significant scatter.  

One possible origin for this scatter 
is of course potential variability with time, and this is precisely what we 
observe for LQ~Hya, with a shear varying from 0.20~\rpd\ at epoch 2000.93 to 
only 0.014 and 0.087~\rpd\ for Stokes $I$ and $V$ data respectively at epoch 
2001.99, i.e.\ within only one year.  Note that this variation of \dom\ as 
measured from Stokes $I$ data is detected at a level of $8.3~\sigma$ while the 
corresponding one in \omeq\ amounts to about $6.6~\sigma$, implying that both 
can be considered as fully reliable.  A similar, though inverse, variation must 
have occurred at least once between epoch 1996.99, at which we detect no 
differential rotation (or at least a value compatible with solid body rotation 
within $2~\sigma$) at an accuracy of 0.024~\rpd, and epoch 2000.93 where this 
very strong shear of 0.2~\rpd\ is observed.  It suggests that the convective 
envelope may indeed be passing cyclically from stages where it rotates almost 
as a solid body, to others where it exhibits strong differential rotation 
between the equator and pole (and likely throughout its depth as well);  we 
discuss this point more extensively in Sect.~\ref{sect:discussion}.   

As for AB~Dor, we also notice a clear difference between the shear as estimated 
from Stokes $I$ and $V$ data.  While the shear measured from unpolarised data 
at epoch 2001.99 is only 0.014~\rpd, the one derived from Zeeman signatures is 
about 6 times larger!  Note that larger shears are again obtained from 
magnetic images, as was already the case for AB~Dor.  No such difference is 
observed at epoch 2000.93;  one must however keep in mind that error bars at 
that epoch are considerably larger than those from our most recent data set.  
We again speculate that this difference most likely reflects the fact that the 
cool spots and magnetic regions we observe in our reconstructed images are 
dissimilar in their essence, and turn out in particular to be anchored at 
different depths in the convective zone.  We must admit that this conclusion 
is very surprising, although not totally unexpected from the fact that only 
very weak spatial correlation was observed between the reconstructed magnetic 
and brightness features (e.g.\ Donati \& Cameron 1997; Donati 1999; Donati et 
al.\ 2003);  the fact that these two classes of photospheric features do not 
witness the same rotational shear strengthens the reality of this difference, 
and may eventually give us more insight on its physical nature.  

\begin{table*}
\caption[]{Same as Table~\ref{tab:abdor} for HR~1099}
\begin{tabular}{ccccccccccc}
\hline
        & \multicolumn{5}{c}{Stokes $I$ data} & \multicolumn{5}{c}{Stokes $V$ data} \\
 &&&&&&&&&& \\
Epoch   & \omeq\ & \dom\ & $\cos^2 \theta_s$ & $\Omega_s$ & $n$ & \omeq\ & \dom\ & $\cos^2 \theta_s$ & $\Omega_s$ & $n$ \\
        & (\mrpd) & (\mrpd) & & (\rpd) & & (\mrpd) & (\mrpd) & & (\rpd) & \\
\hline 
2000.93 &                 &              &       &       &       & $2,222.2\pm4.2$ & $21.2\pm8.1$ & 0.477 & 2.212 & 1,519 \\
2001.99 & $2,223.5\pm2.1$ & $-2.5\pm4.8$ & 0.425 & 2.225 & 6,156 & $2,220.3\pm3.2$ & $-0.4\pm7.1$ & 0.431 & 2.221 & 1,539 \\
\hline 
\end{tabular}
\label{tab:hr1099}
\end{table*}

\subsection{The K1 subgiant of the RS~CVn system HR~1099} 
\label{sect:hr1099}

For HR~1099, only two data set turned out to be of interest for our particular 
purpose;  while those obtained on 1995.94 and 1996.99 are essentially too 
sparse to yield anything useful, that secured at epoch 1999.0 is plagued by 
numerous observations with poor quality (see Donati et al.\ 2003), as was 
already the case for LQ~Hya.  Even Stokes $I$ data from epoch 2000.93 are 
found to produce \chisq\ maps with no clearly defined minimum, but rather 
an infinitely long valley characterising cases where our method fails.  The 
exact reason for which this failure occurs is not entirely clear yet;  all 
conditions for a success of the method, and in particular the presence of 
cool spots on the brightness image at both low and high latitudes as well as 
a relatively long series of observations that regularly come back on the 
same regions of the stellar surface, were however met in this data set.  We 
speculate that it may be due to the fact that some other source of variability 
in the spot distribution, intrinsic in nature (like for instance the formation 
of new spots and the disappearance of old ones), was operating at a much higher 
level at that specific epoch, preventing the signal from the systematic 
evolution induced by differential rotation to build up properly in the \chisq\ 
maps.  The results we obtain are listed in Table~\ref{tab:hr1099} and shown in 
Fig.~\ref{fig:hrdriv}, and were derived using the imaging parameters 
determined by Donati et al.\ (2003), equal to $i=38\degr$ and $\vsin=39$~\kms.   

\begin{figure}
\flushleft{\psfig{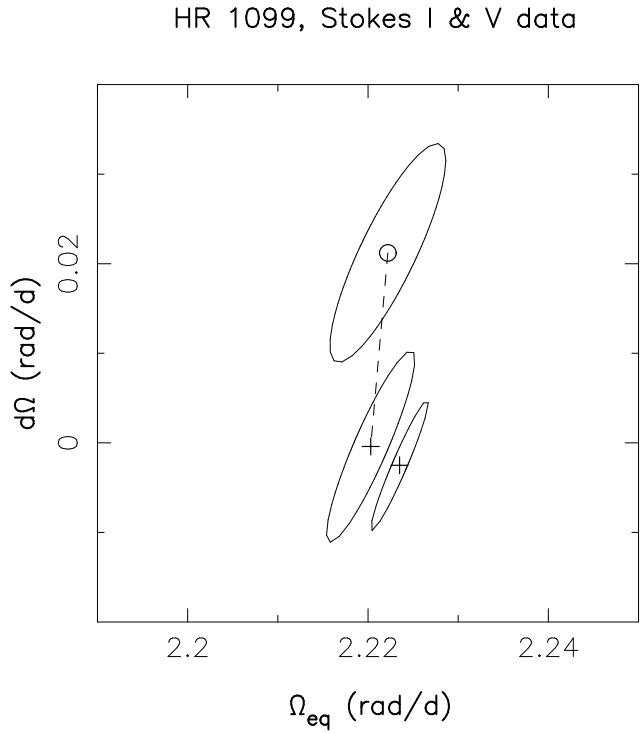}}
\caption[]{Same as Fig.~\ref{fig:lqdriv} for HR~1099.  The point marked with 
$\circ$ now corresponds to epoch 2000.93. } 
\label{fig:hrdriv}
\end{figure}

The error bars we obtain are still too large to claim that HR~1099 is not 
rotating as a solid body;  even at epoch 2000.93 where the signal is strongest, 
the value of \dom\ we derive differs from 0 by no more than $2.6~\sigma$, which 
we do not consider as large enough to be called a reliable detection.  Our 
estimates nevertheless indicate altogether a tendency for the differential 
rotation of HR~1099 to be solar like (i.e.\ with a positive \dom), at least at 
the time of our observations, rather than antisolar (as announced by various 
authors in the recent years, e.g.\ Vogt et al.\ 1999;  Strassemeier \& Bartus  
2000).  Our measurements also pin down the strength of the photospheric shear 
to something of the order of 10~\mrpd, in reasonable agreement to the actual 
estimates obtained by Petit et al.\ (2003a) from independent data sets.  
If we average together the values of \dom\ derived from Stokes $I$ and $V$ 
data at epoch 2001.99, we obtain that the shear is compatible with 0, with 
an error bar of 5~\mrpd.  This is slightly smaller than the value found by 
Petit et al.\ (2003a) at the same epoch (equal to $18\pm4$~\mrpd) and may 
indicate that our error bars are slightly underestimated (as suggested by 
the simulations of Petit et al.\ 2002 in a case with similar phase sampling).  
Since no clear detection is obtained, nothing can of course be quoted on a 
potential variation with time of the differential rotation parameters.

\section{Summary and discussion} 
\label{sect:discussion}

\subsection{Differential rotation, dynamo processes and activity cycles} 

We applied in this paper a new technique for measuring surface differential 
rotation in cool active stars featuring brightness and magnetic spots on their 
surface, and using these surface features as tracers of the stellar rotation 
at various latitudes.  First introduced by Donati et al.\ (2000) then validated 
through extensive simulations by Petit et al.\ (2002), this method is found to 
produce very accurate and reliable results on the three stars we applied it to 
in this new study.  

We obtain positive detections of surface differential rotation in two of our 
three stars (namely the young K0 dwarfs AB~Dor and LQ~Hya).  In the first of 
these two stars, the photospheric shear (and equatorial rotation rate) we 
measure is in very good agreement with earlier estimates published by Donati \& 
Cameron (1997) and Donati et al.\ (1999) from the same data sets, but with a 
different analysis.  In the second object, surface differential rotation is 
detected for the first time, and turns out to be rather large (as much as $3\pm1$ 
times that of the Sun at epoch 2000.93).  In both stars and at all epochs at 
which our method is successful, the 
photospheric shear that we measure is found to be solar like, i.e.\ with the 
stellar equator rotating faster than the pole.  No evidence for anti solar 
differential rotation such as that claimed by, e.g., Vogt et al.\ (1999) or 
Strassemeier \& Bartus (2000) is found in our data.  For the third star (the K1 
subgiant of HR~1099), we only obtain that the photospheric shear (if present) 
is of the order of 10~\mrpd\ (i.e.\ about 5 times weaker than that of the Sun), 
in good agreement with the recent estimates obtained by Petit et al.\ (2003a).  

For both stars on which differential rotation is detected, we also observe 
significant variation of the differential rotation parameters with time.  
This confirms in particular the earlier claim by Cameron \& Donati (2002) that 
the photospheric shear of AB~Dor was fluctuating from epoch to epoch, and 
amplifies the importance and significance of this discovery by revealing that 
other similar stars actually behave in a way roughly identical in essence, 
though much more extreme in intensity.  Our observations indeed disclose that, 
while the temporal variation of the differential rotation parameters remains 
moderate in AB~Dor (smaller than 40~\mrpd\ for \dom\ and 15~\mrpd\ in \omeq\ 
peak-to-peak throughout our 5 epochs of observations), it can be much larger 
in LQ~Hya where it can switch from virtually solid body rotation (e.g.\ 
at epoch 1996.99) to a shear as strong as 200~\mrpd\ with an increase of about 
80~\mrpd\ in \omeq, an effect about 5 times larger in amplitude than what we 
witnessed for AB~Dor.  

Another important discovery reported in this paper is the fact that the 
differential rotation parameters we obtain from Stokes $I$ and Stokes $V$ data 
are not compatible to within the measurement error bars.  We find that this 
effect, although strange, cannot be attributed to a systematic problem in the 
method (see in particular Sect.~\ref{sect:abdor} where this is discussed 
extensively).  We are thus left with the conclusion that this discrepancy 
actually reflects some real property of brightness and magnetic surface 
features of active stars, and therefore of the convective zones in which they 
are anchored.  A possible interpretation of this result (already mentioned 
above) is that the discrepant angular velocity laws respectively derived from 
magnetic and brightness features correspond to different layers within the 
convective zone, e.g.\ to different anchoring depths of brightness and magnetic 
surface regions.  If this is the true explanation, the large difference in 
angular shear witnessed by both types of surface features at a given epoch 
(e.g.\ reaching up to 72~\mrpd\ between the pole and equator on LQ~Hya at 
epoch 2001.99) strongly argues in favour of dynamo processes, and thus 
potential anchoring locations of surface features, being distributed throughout 
the whole convective zone rather than being confined at its base as in the Sun.  
In this context, the observed discrepancy may serve as an indirect tool for 
studying angular velocity fields within stellar convective zones (see 
Sect.~\ref{sect:momentum}).  

The first suspicion is that the temporal variations in differential rotation 
that we detect only reflect the fact that differential rotation is not a 
strict rule that all surface features must obey, but only an average behaviour 
that cool spots and magnetic regions follow in a statistical way;  this is in 
particular what is observed on the Sun (with various classes of differential 
rotation tracers behaving differently and various tracers within each class 
departing from the average differential rotation law for this class), 
reflecting the fact that differential rotation is a consequence of turbulence 
and inevitably contains some randomness when looking at individual features.  
However, the temporal variations we report here (especially in the case of 
LQ~Hya) are so large that we simply cannot invoke the random nature of 
differential rotation to account for the observed variations, and have no 
better explanation to put forward than assuming that the average differential 
rotation law itself is changing as a whole from year to year.  

In this context, a natural idea is that these variations 
are caused by the stellar magnetic cycle converting periodically 
kinetic energy within the convective zone into large scale magnetic fields and 
vice versa, as originally proposed by Applegate (1992) in the particular case of 
close binary systems;  if his scenario is correct, there is of course no reason 
for which all similarly active stars should not exhibit the same phenomenon.  
Although this interpretation would be conceptually very satisfying, it is 
nevertheless premature to conclude already that this is indeed what we 
observe;  a definite demonstration would require for instance to monitor a few 
stars for a long enough time and see both the differential rotation parameters 
and the global polarity of the large scale dynamo field to show cyclical 
variations on the same temporal period, or at least to exhibit strongly 
correlated fluctuations in case of a non-cyclic behaviour.  All activity 
proxies estimated from the reconstructed stellar images (such as the total spot 
coverage or the mean quadratic magnetic flux) and monitored during the last 
decade did exhibit no clear evidence of cyclic variations with time (Donati et 
al.\ 2003a).   

\subsection{Angular momentum and kinetic energy in a differentially rotating convective envelope} 
\label{sect:momentum}

Looking at how \omeq\ varies as a function of \dom\ in any active star for 
which both parameters, as well as their variations, can be estimated can also 
provide us with additional clues on how angular rotation is distributed within 
their convective envelope.  In this initial investigation, we suggest to 
explore this idea by assuming a priori specific angular velocity fields within 
the convective envelope, and derive what relation angular momentum conservation 
imposes between \omeq\ and \dom.  One may argue that angular momentum 
conservation does not apply to the convective zone as a whole since this zone 
regularly exchanges angular momentum both with the circumstellar environment 
(through magnetospheric confinement of massive prominences and winds) and with 
the radiative interior (through a weak magnetic coupling for instance).  These 
external phenomena are however known to occur on a much longer timescale than 
the redistribution of angular momentum within the convective zone itself;  for 
the purpose of this first basic study, we can therefore safely assume that the 
convective zone obeys global conservation of angular momentum.  

The specific angular velocity fields we propose to consider in this paper are 
that of the Sun (with angular rotation independent on the radial distance to 
the centre of the star), and that supposed to occur in very rapid rotators 
(with angular rotation constant along cylinders symmetric with respect to the 
rotation axis).  Of course, we expect many more situations to occur potentially 
in real stars, but the two selected ones likely represent the extreme cases we 
may encounter among all active stars.  
In both cases, we assume that the surface differential 
rotation obeys Eq.~\ref{eq:diffrot}.  We obtain that the internal angular 
velocity field $\Omega(r,\theta)$ is given, in the first of our two test cases, 
by: 
\begin{equation}
\label{eq:drsun}
\Omega(r,\theta) = \ompo + \dom \sin^2 \theta 
\end{equation}
\noindent where $r$ denotes the radial distance to the centre of the star,  
$\theta$ the colatitude with respect to the rotation axis and \ompo\ the 
rotation rate at the pole (equal to $\omeq-\dom$).  In our second test case, 
this equation becomes: 
\begin{equation}
\label{eq:drufr}
\Omega(r,\theta) = \ompo + \dom \left( \frac{r}{R} \right)^2 \sin^2 \theta 
\end{equation}
\noindent with $R$ denoting the stellar radius.  

In this context, we can evaluate the total angular momentum $J$ within the 
convective zone by integrating over all elementary plasma torus of spherical 
coordinate ($r$, $\theta$) and mass $d^2m$ (equal to 
$2\pi\rho(r) r^2\sin\theta\ d\theta dr$ where $\rho(r)$ is the local mass 
density, assumed to depend on $r$ only).  The resulting expression writes: 
\begin{equation}
\label{eq:angmom}
J = 2\ \int_{\rc}^{R} \int_{0}^{\pi/2} r^2 \sin^2 \theta\ \Omega(r,\theta)\ d^2m.   
\end{equation}
\noindent Replacing $\Omega(r,\theta)$ with Eq.~\ref{eq:drsun} and integrating 
over $r$ and $\theta$ yields the following expression for $J$ in our first 
test case: 
\begin{equation}
\label{eq:momsun}
J = \frac{2 \alpha}{3}\ \mc\ R^2\ (\omeq\ - \frac{1}{5}\ \dom )
\end{equation}
\noindent where \mc\ represents the total mass in the convective zone and 
$\alpha$ the normalised mass-weighted second order moment of the fractional 
radius in the convective zone, given by 
$\int_{\rc}^{R} (r/R)^2\ dm / \int_{\rc}^{R} dm$ 
\noindent ($dm$ being the total mass of the elementary plasma shell of 
radius $r$, equal to $4\pi r^2 \rho(r) dr$).  
The same process applied to our second test case yields a similar expression 
for $J$, that writes: 
\begin{equation}
\label{eq:momufr}
J = \frac{2 \alpha}{3}\ \mc\ R^2\ (\omeq\ - \lambda\ \dom )
\end{equation}
\noindent where 
\begin{equation}
\label{eq:lambda}
\lambda = 1-\frac{4}{5}\ \frac{\beta}{\alpha} 
\end{equation}
\noindent and $\beta$ the fourth order moment of the fractional radius 
$\int_{\rc}^{R} (r/R)^4\ dm / \int_{\rc}^{R} dm$.  These parameters mainly 
depend on the detail of the internal stellar structure and are listed in 
Table~\ref{tab:param} for our three stars of interest, using the latest 
available stellar structure models (Charbonnel et al.\ 1996; Siess et 
al.\ 2000).  

\begin{table}
\caption[]{Model parameters depending on the internal structure 
(sequentially the total mass $M$, the external 
radius $R$, the fractional radius at the bottom of the convective zone $\rc/R$, 
the fractional mass included in the convective zone $\mc/M$, the mean surface 
rotation rate $\left< \Omega \right>$, the second, fourth and sixth order moment of the 
fractional radius in the convective zone $\alpha$, $\beta$ and $\gamma$, as 
well as the model parameters $\lambda$ and $\mu$ defined in the text) 
for AB~Dor, LQ~Hya and HR~1099}
\begin{tabular}{llll}
\hline
Parameter     & AB~Dor & LQ~Hya  & HR~1099  \\
\hline 
$M$ (\msun)     & 1.0                & 0.95               & 1.0 \\
$R$ (\rsun)     & 1.05               & 0.95               & 3.7 \\
$\rc/R$         & 0.73               & 0.73               & 0.16 \\
$\mc/M$         & 0.074              & 0.074              & 0.73 \\
$\left< \Omega \right>$ (\ps)  & $1.4\times10^{-4}$ & $4.5\times10^{-5}$ & $2.6\times10^{-5}$ \\ 
\hline 
$\alpha$        & 0.58               & 0.58               & 0.27 \\ 
$\beta$         & 0.35               & 0.35               & 0.11 \\
$\gamma$        & 0.22               & 0.22               & 0.054 \\ 
$\lambda$       & 0.52               & 0.52               & 0.68 \\
$\mu$           & 0.027              & 0.027              & 0.035 \\ 
\hline
\end{tabular}
\label{tab:param}
\end{table}

The conservation of angular momentum within the convective zone therefore 
implies, in our first test case, the following relation between \omeq\ and 
\dom:  
\begin{equation}
\label{eq:rela}
\omeq = \frac{1}{5}\ \dom\ + \omsb, 
\end{equation}
\noindent \omsb\ being a constant, equal to the rotation rate the stellar 
convective zone would have if spinning as a solid body.  In our second test 
case, this relation becomes: 
\begin{equation}
\label{eq:relb}
\omeq = \lambda\ \dom\ + \omsb,
\end{equation}
\noindent implying in particular that the linear relationship between \dom\ 
and \omeq\ is significantly less steep than in our first test case (see 
Table~\ref{tab:param}).  It may be worthwhile to note that, although \omeq\ 
and \dom\ refer specifically to the differential rotation parameters at the 
surface of the star, Eq.~\ref{eq:rela} still holds when replacing \omeq\ and 
\dom\ by their equivalents estimated at radial distance $r$ from the centre 
of the star (and noted respectively $\omeq(r)$ and $\dom(r)$) since both do not 
show any dependence with $r$ in this particular case;  however, this is not 
true in the second test case, where Eq.~\ref{eq:relb} becomes, when expressed 
as a function of $\omeq(r)$ and $\dom(r)$: 
\begin{equation}
\label{eq:relc}
\omeq(r) = \lambda'(r)\ \dom(r)\ + \omsb,
\end{equation}
\noindent where $\lambda'(r)$ is equal to 
\begin{equation}
\label{eq:lambdap}
\lambda'(r) = 1-\frac{4}{5}\ \frac{\beta}{\alpha\ \xi(r)} 
\end{equation}
\noindent with $\xi(r)$ denoting the squared fractional radius $(r/R)^2$.   In 
our second test case, we therefore expect the relationship between $\dom(r)$ 
and $\omeq(r)$ to change (turning steeper with decreasing $r$) depending on the 
depth at which the tracers used to estimate differential rotation are anchored.  

We can also derive by the same method an expression giving the kinetic energy 
$K$ within the convective zone.  The general formula, writing: 
\begin{equation}
\label{eq:kinen}
K = \int_{\rc}^{R} \int_{0}^{\pi/2} r^2 \sin^2 \theta\ \Omega(r,\theta)^2\ d^2m, 
\end{equation}
\noindent becomes, in our first test case: 
\begin{equation}
\label{eq:kesun}
K = \frac{\alpha}{3}\ \mc\ R^2\ \left[ (\omeq\ - \frac{4}{5}\ \dom )^2 + \frac{8}{175}\ \dom^2 \right] 
\end{equation}
and in the second test case: 
\begin{equation}
\label{eq:keufr}
K = \frac{\alpha}{3}\ \mc\ R^2\ \left[ (\omeq\ - \lambda\ \dom )^2 + \mu\ \dom^2 \right] 
\end{equation}
\noindent where $\mu$ stands for: 
\begin{equation}
\label{eq:mu}
\mu = \frac{24}{35}\ \frac{\gamma}{\alpha}\ - \frac{16}{25}\ \left( \frac{\beta}{\alpha} \right)^2
\end{equation}
\noindent and $\gamma$ for the sixth order moment of the fractional radius 
$\int_{\rc}^{R} (r/R)^6\ dm / \int_{\rc}^{R} dm$.  In both cases, we obtain 
that $K$ can be written as the sum of two terms, one that we know is constant 
(given Eqs.~\ref{eq:rela} and \ref{eq:relb}, reflecting the conservation of 
angular momentum) and a second, potentially variable, one, directly 
proportional to $\dom^2$.  To compute the actual variation in kinetic energy 
$dK$, one must note that \dom\ can itself be written as the sum of two terms: 
\begin{equation}
\label{eq:dom}
\dom = \dommin\ + \domvar 
\end{equation}
\noindent where \dommin\ and \domvar\ respectively represent the minimum value 
of the observed photospheric shear (that corresponding to the state closest to 
solid body rotation) and the variable contribution to \dom.  In this context, 
the variation of kinetic energy induced by the differential rotation changes 
writes:  
\begin{equation}
\label{eq:dk}
dK = \frac{\alpha}{3}\ \mc\ R^2\ \mu\ ( \domvar^2\ + 2\ \domvar\ \dommin) 
\end{equation}
\noindent ($\mu$ being replaced by 8/175 in our first test case).  

It could be tempting to consider that the whole variation in kinetic energy is 
transformed into magnetic energy as part of the magnetic cycle (as proposed by 
Lanza et al.\ 1998).  However, this conclusion first requires that we obtain 
an expression similar to that derived above for $dK$ (see Eq.~\ref{eq:dk}), 
but for the associated changes in potential energy that result from the 
fluctuations in the internal angular velocity field and thus to the variations 
in the shape of the star.  We could then derive which fraction of the kinetic 
plus potential energy is converted into magnetic energy during the activity 
cycle.  Similarly, we could also obtain estimates of the fluctuations in the 
in the stellar quadrupole moment (as done for instance by Applegate 1992), but 
in the specific context of our study (i.e.\ with the internal velocity fields 
assumed above in Eqs.~\ref{eq:drsun} and \ref{eq:drufr}).  We however keep this out 
of this initial study and postpone it for future papers;  it is nevertheless 
rather safe to claim that the amount of kinetic energy transformed into 
magnetic energy is of the same magnitude as (though not strictly equal to) 
$dK$.  

\subsection{Internal velocity fields and magnetic energy fluctuations of active stars} 
\label{sect:velocity}

If we now compare the model predictions obtained in Sect.~\ref{sect:momentum} 
with our observations, we straightforwardly obtain that the second test case 
(featuring constant angular velocity along cylinders within the convective 
zone) seems much more appropriate.  It is indeed only in this case that 
different depths in the convective zone correspond to different angular shears 
(see Eqs.~\ref{eq:drsun} and \ref{eq:drufr}), thus providing a natural explanation 
of the discrepancy between the \omeq\ and \dom\ estimates derived from Stokes 
$I$ and $V$ data.  If this is indeed the correct interpretation, it would imply 
that magnetic features, showing the strongest shear, are in general anchored  
closer to the surface (where $\dom(r)$ is maximum) than cool spots.  This is 
at least in qualitative agreement with the conclusions of, e.g., Donati et al.\ 
(2003), claiming that magnetic features (and in particular those hosting 
predominantly azimuthal fields) must be formed very close to the surface in 
these stars.  Further support about the idea that dynamo magnetic fields (and 
in particular their toroidal components) can be produced very close to the 
stellar surface comes from the recent discovery that the Sun features a 
subphotospheric layer with a strong radial gradient in angular rotation 
(Corbart \& Thompson 2002) that seems at least able to trigger dynamo action 
(Dikpati et al.\ 2002).  

However, there is already a number of potential problems with 
this simple description.  It implies for instance that the range of possible 
latitudinal shears we should observe on a given star at a given epoch goes from 
\dom\ at the surface, down to $\dom(\rc)=\dom\ \xi(\rc)$ at the bottom of the 
convective zone, i.e.\ that the maximum ratio between the shears we measure 
from Stokes $V$ and Stokes $I$ data should be $1/\xi(\rc)$.  Although this is 
verified for AB~Dor (for which the measured ratio varies from 1.25 to 1.8 and 
thus remains always smaller than the maximum predicted value of $1/0.73^2=1.9$), 
the situation is different for LQ~Hya where we observe (at epoch 2001.99) 
differential rotation shears in Stokes $V$ and Stokes $I$ that differ by a 
factor of about 6 (see Table~\ref{tab:lqhya}), much larger than the theoretical 
maximum of 1.9.  The origin of this discrepancy is not clear yet;  the fact that
only one point (out of the seven for which measurements from Stokes $I$ and $V$ 
data are available) exhibits this problem tells at least that the model we 
propose, although probably oversimplified, is not necessarily a complete 
nonsense.  One possibility is that the model we consider does not take 
into account the additional shears that probably exist in the interface layer 
between the convective envelope and the radiative interior, nor that associated 
with a potential subsurface shear similar to that of the Sun (Corbard \& 
Thompson 2002);  we may thus imagine that, in some rare cases, the bulk of 
the differential rotation tracers we use is anchored in these thin shear layers, 
producing a small number of observations in disagreement with our simple picture.  
It may also indicate that the differential rotation estimate secured for LQ~Hya 
at epoch 2001.99 from Stokes $I$ data is affected by some spurious contribution 
that we have not yet identified (values derived from Stokes $V$ data being 
always more reliable and less prone to systematic effects);  although we are of 
course fully aware that is a rather unsatisfying and potentially worrying 
statement, we cannot yet rule out this possibility completely until a larger 
sample of estimates are available for us to check if this problem is truly an 
exception or rather a regular occurence.  In the rest of the study, we 
exclude this point from the analysis and further investigate how compatible is 
the toy model we propose with the rest of our data.  

A quantitative test consists in measuring the slope of the \dom\ versus 
\omeq\ relations that we obtain for each star by taking into account all 
estimates derived in Sect.~\ref{sect:diffrot} (and shown on 
Figs.~\ref{fig:abdriv} to \ref{fig:hrdriv}) and comparing it to those the 
model predict (see Eqs.~\ref{eq:rela} to \ref{eq:relc}).  To determine the 
optimal linear relationship that provides the best fit through all data 
points and their associated confidence ellipses for each star (something 
that a conventional least-squares fit, not taking into account the 
correlated errors on both axes, cannot properly achieve), we iteratively 
consider a large number of such relationships (i.e.\ a large number of  
$\lambda$ and \omsb\ pairs) and compute, for each of them, the closest approach 
(in terms of \chisq) to all error ellipses and the associated overall goodness 
of fit (defined as the sum of the minimum square distances to each ellipse). 
From these values, we derive a \chisq\ map (in the $\omsb-\lambda$ plane), 
look for a minimum 
and fit the map by a bi-dimensional paraboloid in the neighbourhood of this 
minimum (by a method similar to that described in Sect.~\ref{sect:modelling}) 
to obtain both the optimum values of $\lambda$ and \omsb\ as well as the 
corresponding error bars.  

For AB~Dor and using all the estimates listed in Table~\ref{tab:abdor} 
(10 points altogether with their associated error ellipses), we obtain from 
this procedure that the minimum achievable reduced \chisq\ is 10.5, already 
indicating that the model provides a rather poor explanation of our 
observations.  When 2 (out of the 10 available) differential rotation 
estimates are removed from the full sample (those obtained from Stokes $I$ 
data at epoch 1995.94 and from Stokes $V$ data at epoch 2001.99) on the 
reason that they deviate most from the fit including all points (at a level 
of about 5~$\sigma$), the minimum achievable reduced \chisq\ decreases to 
a smaller, but still uncomfortably high, value of 4.8, yielding values 
of $\lambda$ and \omsb\ equal to $0.48\pm0.08$ and $12.212\pm0.005$~\rpd\ 
respectively (with 1~$\sigma$ error bars enhanced by a factor of 
$2.2=(4.8)^{1/2}$ to reflect the poor fit obtained with this model).  
Note that the value of $\lambda$ we derive agrees well with that 
predicted for the second test model (equal to 0.52, see Table~\ref{tab:abdor}).  
If we now impose $\lambda$ to be equal to 0.2 (the value predicted for our 
first test model), the minimum reduced \chisq\ that we can now achieve is 
about 2.6 times larger than that obtained without constraining $\lambda$.  
This strongly suggests that our first model (assuming solar-like differential 
rotation within the convective zone) is a much less probable option than the 
second one (assuming constant angular velocity along cylinders within the 
convective zone), event hough the accuracy to which the observations are 
fitted with this second model is still not satisfactory.  

To investigate whether the agreement with the data can be improved, the 
best way is to attempt exploiting all built-in characteristics of our 
simple model, for instance by taking explicitly into account, when 
comparing with the data, the fact that the different types of tracers may 
be anchored at different convective depths.  To achieve this, we proceed 
along the lines presented in Sect.~\ref{sect:momentum} and try to measure 
the differential rotation estimates corresponding to Stokes $I$ and 
Stokes $V$ data with different linear relationships (using 
Eqs.~\ref{eq:relc} and \ref{eq:lambdap}).  More specifically, we proceed 
as explained in the paragraphs above, except that we now have three 
independent parameters, $\lambda$ (the slope of the \omeq\ versus \dom\ 
relationship in the Stokes $V$ data, presumably referring to the surface 
layers), $\lambda'$ (the slope of this relationship in the Stokes $I$ 
data, presumably referring to deeper convective layers) and one value of 
\omsb.  The only difference from the previous procedure is that we now fit 
a three dimensional (instead of a two dimensional) paraboloid to the \chisq\ 
map.  This new modelling attempt (with the same two data points excluded) 
yields a minimum reduced \chisq\ of 3.2, along with values of $\lambda$, 
$\lambda'$ and \omsb\ respectively equal to $0.34\pm0.06$, $0.28\pm0.08$ 
and $12.222\pm0.004$~\rpd\ (error bars being again scaled up in proportion 
to the square root of the reduced \chisq).  Although the fit to the data 
is now more satisfactory and the derived value of $\lambda'$ falls within 
the expected range (between 0.13 and 0.40, given the inverse squared 
fractional anchoring depths $1/\xi$ obtained from ratioing the estimated 
shears at each epoch, see above), the value of $\lambda$ we extract 
is poorly compatible with that we expect for the surface layers (equal to 
0.52), with a mismatch of about 3~$\sigma$.  Repeating the procedure 
with $\lambda$ being fixed to the expected value of 0.5 yields no 
improvement in the accuracy level to which the data are fitted with 
respect to the two parameter fit described above.  

We therefore conclude that the estimates of differential rotation and of 
its temporal variations at the surface of AB~Dor suggest that the 
distribution of angular velocities within the convective zone is closer to 
that we expect for rapid rotators than to that of the Sun.  However, the 
simple model we propose, in addition to the fact that the accuracy level to 
which it fits the data is only rough, also predicts some features that are 
not reproduced in the observations, and in particular the different slopes 
of the linear \omeq\ versus \dom\ relationships expected for tracers 
anchored at different convective depths.  What it may simply indicate is 
that our test models are still far too simple to match the observations at 
that level of detail.  Further attempts, with more realistic large scale 
angular velocity fields in the convective zone will be the subject of 
forthcoming papers.  

The first part of this experiment can also be applied to our LQ~Hya data.  
Using the two parameter fitting procedure described above through the 
available measurements (4 points with associated error ellipses, once 
discarded the estimate from Stokes $I$ data obtained at epoch 2001.99, 
see above), we obtain values of $\lambda$ and \omsb\ that are respectively 
equal to $0.36\pm0.03$ and $3.912\pm0.003$~\rpd\ and provide a 
nice fit to the data, with an associated reduced \chisq\ level of 0.84.  
For this star again, we obtain that the slope of the observed \omeq\
versus \dom\ linear relationship is not compatible with a solar like 
differential rotation within the convective zone;  repeating the same 
procedure with $\lambda$ fixed to 0.2 (as required by our first test 
model) increases the minimum reduced \chisq\ up to 7.2, i.e.\ 8.4 times 
that achieved with the unconstrained fit.  
Assuming now constant angular velocity along cylinders, the model would 
then require the tracers to be anchored at about mid depth within the 
convective zone (at a fractional radius of about 0.86) to agree with 
the observations.  Another possibility is of course that the internal 
velocity field of LQ~Hya is somehow intermediate between that of a very 
fast rotator like AB~Dor and that of a slow rotator like the Sun, 
reflecting the fact that the rotation rate of LQ~Hya (3.9~\rpd) is 
bracketed by that of AB~Dor (12.2~d) and that of the Sun (0.25~\rpd).  
In this context, contours of constant angular rotation within the 
convective zone should start deviating from being aligned with the rotation 
axis, and get slightly tilted towards the radial direction.  Once again, 
we caution that these conclusions are still highly speculative and require 
more observational material (in particular measurements on other stars 
rotating slower and faster than LQ~Hya) to be settled more firmly.  

Carrying out a similar investigation in the particular case of HR~1099 would 
also be very interesting, as our second model predicts a significantly higher 
slope for the \omeq\ versus \dom\ linear relation (equal to about 0.68, see 
Table~\ref{tab:param}) due to the fact that the convective zone (pertaining 
more than 99\% of the total stellar volume) is much deeper in this star.  The 
smaller rotation rate of HR~1099 (2.2~\rpd) may also act at the same time to 
decrease the value of $\lambda$ with respect to the predicted one.  
Since no temporal variations of 
differential rotation are however yet detected on HR~1099, all this remains 
essentially a matter of speculation, until more accurate estimates (such as 
those of Petit et al.\ 2002, but covering a longer time span) are available.  

Another important aspect of this problem concerns the amount of kinetic energy 
that the star either withdraws from or releases into the total energy reservoir 
of the convective zone when switching from one state of differential rotation 
to another.  In particular, we should make sure that the total power associated 
with the energy transfer from or into kinetic energy is significantly smaller, or 
at the very least not larger, than the stellar luminosity itself, to ensure 
that the process we invoke to explain our observations (which in principle 
should apply to the vast majority of rapidly rotating cool stars) does not 
implicitly require at the same time that the star looses its energy at a very 
fast rate.  Of course, we reckon that the energy transfers we invoke likely 
correspond to some redistribution process between the various reservoirs in 
which the convective zone can store energy (i.e.\ kinetic, potential and 
magnetic energy wells);  however, we suspect that these transfers are  
inevitably associated with some energy dissipation, i.e.\ that a significant 
fraction (the exact proportion being poorly known) of the energy transfered 
is lost (and thus radiated) in the process.  

The changes in kinetic energy associated with the differential rotation 
fluctuations can easily be estimated from Eq.~\ref{eq:dk}.  In the case of 
AB~Dor, the results reported in Sect.~\ref{sect:abdor} suggest that \dommin\ 
and \domvar\ are respectively equal to 60 and 40~\mrpd, implying that $dK$ 
should be equal to $4\alpha \mu\ \mc R^2 \domvar^2/3$ or about $4\times10^{39}$~erg.  
Assuming this transfer occurs progressively over a typical timespan of a 
few years ($10^8$~s), we obtain that the power required to operate the change 
we invoke corresponds to about 1\% of the total stellar luminosity.  (Note that 
$dK$ is only weakly dependent on the assumed internal velocity field, varying 
only by a factor of about two between our two test models).  For 
LQ~Hya, our observations indicate that \dommin\ is probably close to zero 
while \domvar\ gets as large as 200~\mrpd, implying that $dK$ should now be 
equal to $\alpha \mu\ \mc R^2 \domvar^2/3$ or about $2\times10^{40}$~erg.  
If we again consider that the corresponding evolution occurs on a time scale of 
a few years, we find that the associated power corresponds to 10\% of the total 
stellar luminosity.  Although this sounds rather large already, one must keep 
in mind that only a small fraction of this amount is actually lost (i.e.\ 
radiated) in the process;  it suggests that, even in the case of the star that 
exhibits the most drastic effect (i.e.\ LQ~Hya), the corresponding power 
dissipated in the process does probably not exceed 1\% of the stellar 
luminosity.  Since no temporal changes were detected in the case of HR~1099, we 
thus cannot obtain a similar estimate without any (hopefully reasonable) 
assumption on \dommin\ and \domvar.  Taking $\dommin=0$ and $\domvar=40$~\mrpd\ 
as a guess (and to remain grossly compatible with the differential rotation 
changes that the orbital period fluctuations observed on this system request, 
Donati 1999), we obtain that $dK$ reaches as much as $7\times10^{40}$~erg and 
that the associated power required to drive this change corresponds to about 
3\% of the stellar luminosity (provided the change in \dom\ occurs over a 
few years).  

If we finally assume that most of the changes in kinetic energy associated with 
the fluctuations in differential rotation are associated with equal (and 
opposite) variations in magnetic energy occurring throughout the whole 
convective zone, we can obtain a very rough estimate of the average magnetic 
field strength required to drive these changes.  We caution that this estimate 
is necessarily very rough for at least several reasons.  The first reason is of 
course that the magnetic field is expected to be very intermittent throughout 
the convective zone and to exhibit a strong gradient in the radial direction 
(comparable to the one that the gas pressure undergoes), implying that an 
average magnetic field strength over the whole convective volume is not 
extremely meaningful;  the second reason (already mentioned above) is that 
fluctuations in potential energy are also expected to occur (as a consequence 
of the changes in the shape of the star) and probably participate as well in 
the total energy budget.  However, a rough estimate being better than no 
estimate, we decided to mention in the paper the derived average magnetic 
strengths, to at least convince the readers that the magnitude of these fields 
is physically plausible.  For AB~Dor and LQ~Hya, we obtain magnetic intensities 
of 9 and 23~kG respectively;  using the above guesses in the case of HR~1099 
yields a field strength of 5~kG.  This is clearly larger that the magnetic 
strengths estimated at the surface (e.g.\ Donati 1999; Donati et al.\ 2003),  
and reflects the strong radial gradient of field strength (pointing inwards) 
that probably exists within the convective zone.

\section{Conclusion and prospectives}
\label{sect:conclusion}

This paper reports positive detections of surface differential rotation on two 
rapidly rotating cool stars at several epochs, by using stellar surface 
features (both cool spots and magnetic regions) as tracers of the large scale 
latitudinal shear that distorts the convective envelope in this type of stars.  
We also report definite evidence that this differential rotation is different 
when estimated from cool spots or magnetic regions, and that it undergoes 
temporal fluctuations of potentially large amplitude on a time scale of a few 
years.  

We consider these results as further evidence that the dynamo processes 
operating in these stars are distributed throughout the convective zone rather 
than being confined at its base as in the Sun.  With the help of two very 
simple models of the angular velocity field within the convective zone (one 
that resembles that of the Sun and one supposed to represent that of ultra 
fast rotators), we explore the quantitative implications that angular momentum 
conservation within the convective zone imposes on potential variations of 
surface differential rotation with time.  

This modelling suggests, from comparison with our observations, 
that the rotation velocity field within the convective zone of the stars we 
investigated is not like that of the Sun, and may resemble that we expect 
for rapid rotators, with constant angular velocity along cylinders aligned 
with the rotation axis.  The power required to drive the observed changes in 
the differential rotation of the convective zone are estimated to be of the 
order of 1 to 10\% of the total stellar luminosity;  we speculate that these 
changes result from the dynamo processes (and from the underlying magnetic 
cycle) that periodically converts magnetic energy into kinetic energy and vice 
versa.  

The technique outlined in this paper corresponds to the first practical method 
for investigating the large scale rotation velocity field in convective zones 
of cool active stars.  In particular, we note that it can already provide us 
with more information than what one could obtain with other existing techniques 
such as asteroseismology (expected to yield rather poor resolution on the internal 
rotation profile in the radial direction, even when operated from space, at least 
in the next decade).  We therefore strongly argue in favour of renewing the 
experiment reported here, in order to explore more extensively the full 
capabilities of this new method.  Monitoring continuously the same star for time 
spans as long as several months every year for several years would be especially 
interesting if one wants to disclose short term chaotic variability (presumably 
due to the natural randomness inherent to turbulence and differential rotation) 
from the long term supposedly periodic evolution driven by the magnetic cycle.  

In this context, the new spectropolarimeter ESPaDOnS, designed for maximum 
efficiency and being built for the 3.6~m Canada-France-Hawaii Telescope 
(commissioning scheduled for spring 2003), represents the optimal instrumental 
facility for carrying the observations that this new method requires.  We 
expect in particular that the large gain in sensitivity that this new tool 
will provide over older generation instruments should give us the opportunity 
of both extending the available data set to a much larger sample of stars and 
studying the physical processes of interest to us in much more details.  This 
research program will also strongly benefit from having similar instruments 
on smaller telescopes to allow the long, repeated and if possible multisite, 
observational campaigns that are necessary for optimising the performances of 
the investigation we suggest.  In this respect, the 2~m Bernard Lyot Telescope 
atop Pic du Midi in France, that will soon be equipped with NARVAL (a carbon 
copy of ESPaDOnS, whose construction just started) and dedicated to 
spectropolarimetric studies, should be extremely useful for this kind of 
research.

\section*{ACKNOWLEDGEMENTS}

We warmly thank Corinne Charbonnel for kindly providing us with some models of 
the internal structure of the three stars on which this study is focused.    
We are also grateful to the referee, Rob Jefferies, for making very detailed 
and relevant comments that allowed us to improve the paper substantially.

\end{document}